\documentclass[journal]{IEEEtran}
\usepackage{color}
\usepackage{amsmath}

\usepackage{graphics,graphicx}

\usepackage[numbers,sort&compress]{natbib}

\ifCLASSINFOpdf
\else
\fi

\begin{document}

\title{Detection of Orbital Angular Momentum with Metasurface at Microwave Band}
%
%
%

\author{Menglin~L.~N.~Chen,~\IEEEmembership{Student~Member,~IEEE,}
        Li~Jun~Jiang,~\IEEEmembership{Senior~Member,~IEEE,}
        and~Wei~E.~I.~Sha,~\IEEEmembership{Senior~Member,~IEEE}
\thanks{M. L. N.~Chen and L. J. Jiang are with the Department
of Electrical and Electronic Engineering, The University of Hong Kong, Hong Kong (e-mail: menglin@connect.hku.hk; jianglj@hku.hk).}
\thanks{W.~E. I.~Sha is with College of Information Science and Electronic Engineering, Zhejiang University, Hangzhou, 310027, P. R. China. He is on leave from the Department
of Electrical and Electronic Engineering, The University of Hong Kong, Hong Kong (email: weisha@zju.edu.cn).}}

%
%

\markboth{Ieee antennas and wireless propagation letters,~Vol.~XX, 2017}%
{Shell \MakeLowercase{\textit{et al.}}: Bare Demo of IEEEtran.cls for IEEE Journals}
%



\maketitle

\begin{abstract}
An orbital angular momentum (OAM) detection approach at microwave band is proposed. A transmittance function is exploited to model a transmissive metasurface. Then the metasurface is designed to convert an OAM wave to multiple waves, only one of which is gaussian. The radiation direction of the gaussian wave is distinguishable according to the order of incident OAM. Consequently, by locating the gaussian wave, the incident OAM can be conveniently determined. We use a simple field source to simulate the incident OAM wave in full-wave simulation. It largely simplifies the simulation process when an incident wave carrying OAM is needed. Both numerical and full-wave simulation results are provided to validate our design and they show a good agreement with each other. Then, the metasurface is optimized for high directivity. Our work can provide an efficient and effective way for OAM detection in radio communications.
\end{abstract}


\section{Introduction}

\IEEEPARstart{E}{lectromagnetic} (EM) waves can carry orbital angular momentum (OAM), which provides them with an extra degree of freedom. The OAM waves have been applied in wave multiplexing and demultiplexing in communications~\cite{yanyan,wangjian,thide} and microwave imaging~\cite{kangliu}. Recently, research on the generation of OAM has been carried out extensively. Approaches for OAM generation include using spiral phase plates~\cite{spp,zhangxianmin_spp}, antenna arrays~\cite{array,array2}, computer generated holograms~\cite{CGH}, and metasurfaces~\cite{lilong_meta,menglin2}. As a reciprocal process, detection of OAM at the receiver side is of equal importance but challenging due to the divergence and spatial-dependence nature of an OAM wave. The approaches for OAM detection include mode analysis based on field data~\cite{radiosci, zhanglocal} and observation of the rotational Doppler shift~\cite{doppler}. The former approach needs to acquire the distributed complex field data, which is usually realized by near-field scanning technique. The complexity of measurement is high. The latter approach requires the rotation of the OAM wave, so mechanical rotation of the OAM source is needed. 

\begin{figure}[!t]
\centering
\includegraphics[width=\columnwidth]{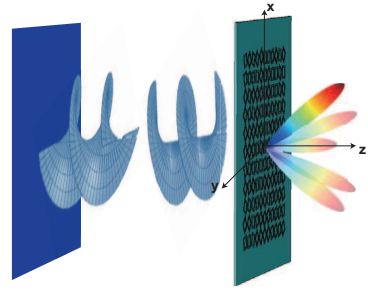}
\caption{Schematic representation of multiple OAM-beam detection by a single metasurface.}
\label{layout}
\end{figure}

It is known that a metasurface can be designed to control EM waves by abrupt phase changes at patterned scatterers on it~\cite{capasso,cuitj_meta,caloz_meta}. The high feasibility of tailoring the geometry of the scatterers makes metasurface a wonderful candidate to achieve various wave manipulation. In this paper, we propose a novel and effective approach for multiple OAM-beam detection by a single transmissive metasurface. This process is schematically summarized in Fig.~\ref{layout}. The wave carrying OAM impinges on the metasurface and then, is converted to a directional beam without OAM. The directivity of the output beam depends on the incident OAM. Furthermore, the metasurface is optimized based on a modified transmittance function to obtain high directivity.

\section{Methodology}
EM wave with OAM possesses a spatial phase dependence expressed by $e^{jl\phi}$, where $l$ is the OAM index and $\phi$ is the azimuthal angle. To generate an OAM of order $l$, scatters on a metasurface with azimuthal location $\phi$ are designed to provide a phase change of $l\phi$.

For the generation of multiple OAM beams, the transmittance of  the metasurface is given by

\begin{equation}
t(r,\phi) = \sum_{m} A_m e^{j(l_m\phi+k_{xm}x+k_{ym}y)},
\end{equation}
where $r$ is the radial position, $\phi$ is the azimuthal position, $A_m$ is the weight of the $m$th beam, $l_m$ is the corresponding OAM index, and $k_{xm}$, $k_{ym}$ are the transverse wave numbers of the $m$th beam.

Then, the far-field response of the metasurface illuminated by an incident wave $E_{in}$ is calculated by doing the Fourier transform~\cite{bookFT}
\begin{equation}
E = F \{E_{in} \cdot t\}.
\end{equation}

If we consider a Laguerre-Gaussian beam (LG$_{pl}$) at normal incidence with radial index $p=0$ and azimuthal index $l_0$, at the focal plane, the beam is expressed by~\cite{bookstructure}
\begin{equation}
E_{in}(r,\phi) = r^{l_0}e^{-r^2/w^2}e^{jl_0\phi},
\end{equation}
where $w$ is the beam waist. 

\begin{figure}[!t]
\centering
\includegraphics[width=\columnwidth]{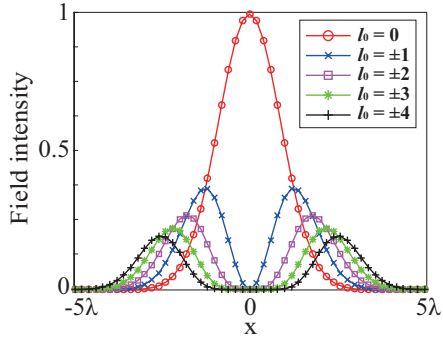}
\caption{Intensity of the electric field along $x$ axis at the focal plane for LG$_{0l_0}$. Each beam carries the same power and the beam waist $w=1.785\lambda$, where $\lambda$ is the free-space wavelength.}
\label{field_line}
\end{figure}

Clearly, the field modulus described in Eq.~(3) is cylindrically symmetric. By setting the same beam power and beam waist for each LG$_{0l_0}$, the field intensity along $x$ axis at the focal plane when $l_0=0,\pm 1, \pm 2, \pm 3, \pm 4$ is shown in Fig.~\ref{field_line}. The field intensity at the beam axis is zero when $l_0 \neq 0$, which is a common feature of an OAM beam. Meanwhile, the beam diverges when $l_0 \neq 0$ and as $|l_0|$ increases, it diverges faster, resulting in a decreased field intensity.

Then, under the illumination of LG$_{0l_0}$, the far-field response is estimated by~\cite{formula}
\begin{align}
\begin{split}
E &= \sum_{m} A_m   F \{r^{l_0}e^{-r^2/w^2} e^{j(l_m+l_0)\phi+j(k_{xm}x+k_{ym}y)} \}\\
&= \sum_{m} A_m F \{ E_{\mathrm{OAM}(l_m+l_0)}(k_{xm}, k_{ym})\}.
\end{split}
\end{align}

Therefore, multiple beams are generated and at the designed k-space position $(k_{xm}, k_{ym})$, the OAM order is $l_m+l_0$ . It can also be noticed in Eq.~(4) that the power of each produced beam, $P_m$ is modulated by $A_m$ and $P_m \propto E_m^2 \propto A_m^2$. In the following, we will set $A_m=1$ so that the generated beams have the same power level.

\begin{figure}[!t]
\centering
\includegraphics[width=\columnwidth]{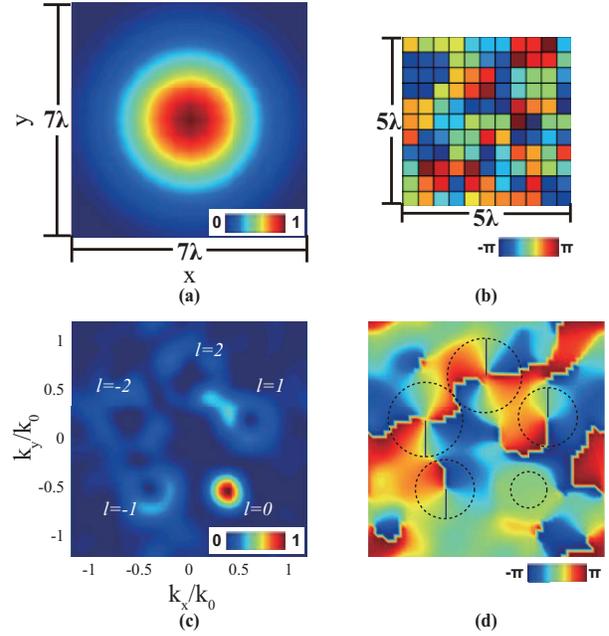}
\caption{Responds of the metasurface under the Gaussian beam incidence. (a) Amplitude of the incident wave with $l_0=0$, $w=30$~mm; (b) phase distribution on the metasurface; (c) the far-field intensity pattern of the diffracted beam; (d) the far-field phase pattern of the diffracted beam.}
\label{matlab_field}
\end{figure}

We design five beams with $l_m=2, 1, 0, -1, -2$ at the directions of $\theta=40^\circ$ and $\phi=90^\circ, 18^\circ, 306^\circ, 234^\circ, 162^\circ$. The corresponding wave numbers are $k_x / k_0 =0, 0.6113, 0.3778, -0.3778, -0.6113$ and $k_y / k_0 =0.6428, 0.1986, -0.5200, -0.5200, 0.1986$, where $k_0$ is the free-space wave number. Then, the transmittance $t(r,\phi)$ is calculated based on Eq.~(1) and the phase information is extracted. The phase distribution is discretized by $11 \times 11$ pixels as shown in Fig.~\ref{matlab_field}(b) so that a metasurface with $11 \times 11$ unit cells can be used for the implementation. Under the Gaussian beam incidence ($l_0=0$), five beams with OAM of order $l_m$ are generated, which is shown in Fig.~\ref{matlab_field}. Since the fives beams are designed to carry the same power, their field intensity follows the trend in Fig.~\ref{field_line}, i.e., as the OAM increases, it decreases. The straight line in the area of each beam in Fig.~\ref{matlab_field}(d) is the first zero-phase line. We find some distortions in the generated beams. There are several reasons: a) Based on the Fourier transform, the larger the metasurface aperture, the more concentrated the generated beam will be. Due to the finite metasurface aperture in real case, the beam will diverge and interfere with each other; b) The divergence nature of the OAM beams will make the interference effect more severe and the larger the OAM is, the more severe the scenario will become; c) The metasurface only retains the phase information in the transmittance function. Therefore, the quality of the generated beam will inevitably be degraded.

\begin{figure}[!t]
\centering
\includegraphics[width=\columnwidth]{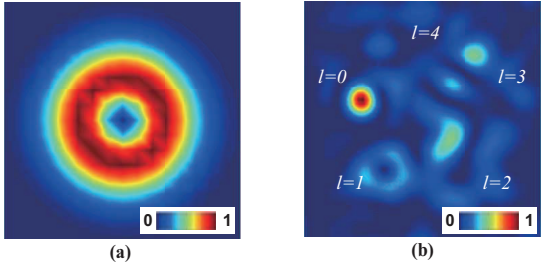}
\caption{Responds of the metasurface under the LG$_{02}$ beam incidence. (a) Amplitude of the incident wave with $l_0=2$, $w=20$~mm; (b) the far-field intensity pattern of the diffracted beam.}
\label{matlab_field2}
\end{figure}

When the incident wave carries an OAM ($l_0$ is chosen from $0, \pm 1, \pm 2$), it is converted to five beams, only one of which is Gaussian beam when $l_0=-l_M$. The output Gaussian beam is at the k-space position $(k_{xM}$, $k_{yM})$. Obviously, for different $l_0$, $M$ is different, so as $(k_{xM}$, $k_{yM})$. Therefore, by examining the field along the five beam axes, we can tell where the Gaussian beam locates, i.e. determine $M$. Then the OAM of the incident beam can be determined. An example with $l_0=2$ is shown in Fig.~\ref{matlab_field2}. As expected, the location of the Gaussian beam is at $(k_{x5},k_{y5})$ because $l_0=-l_5$. This process is reversed when $l_0$ is unknown. 

\section{Simulation}
\begin{figure}[!t]
\centering
\includegraphics[width=\columnwidth]{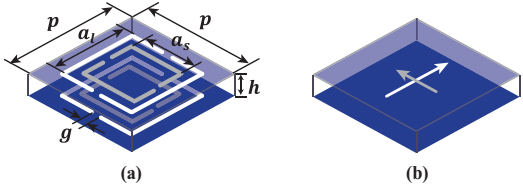}
\caption{Schematic of the unit cell and its equivalent magnetic dipole model. (a) The bilayer complementary unit cell. Each layer contains four split rings. The parameters are: $p=7$~mm, $a_l=5.2$~mm, $a_s=3.9$~mm, $g=0.2$~mm, $h=0.8$~mm and the width of the slots is $0.2$~mm. Substrate is F4B220 with permittivity $\epsilon_r=2.2$ and loss tangent $0.003$; (b) the unit cell is modeled by two orthogonal magnetic dipoles.}
\label{unitcell}
\end{figure}

We use the unit cell proposed in~\cite{menglin1} for the metasurface design. Its geometry and equivalent dipole model are shown in Fig.~\ref{unitcell} and at the designed frequency of $17.85$~GHz, it converts a right (left) circularly polarized wave to a left (right) circularly polarized wave. By axially rotating the unit cell an angle $\alpha$, an additional phase $e^{\pm 2 i \alpha}$ can be introduced. This phase is used to construct the required transmittance function. 

\begin{figure}[!t]
\centering
\includegraphics[width=\columnwidth]{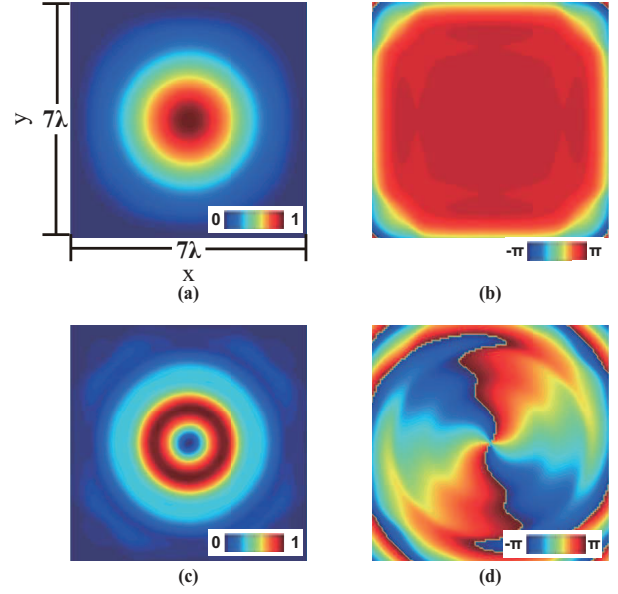}
\caption{Right circularly polarized component of the incident wave at a transverse plane $z=-5$~mm. The focal plane of the wave is at $z=0$. The plotting frequency is $17.85$~GHz. Beam waist is $30$~mm. (a) Amplitude and (b) phase of the incident wave with $l=0$; (c) amplitude and (d) phase of the incident wave with $l=2$.}
\label{cst_field}
\end{figure}

Full-wave simulation is done in CST MWS. Circularly polarized Laguerre-Gaussian sources are required in the full-wave simulation. We import the transverse components $E_x, E_y, H_x, H_y$ of the right circularly polarized $LG_{0l}$ wave on a transverse plane as a field source. Field components of two incident waves $LG_{00}$ and $LG_{02}$ are shown in Fig.~\ref{cst_field}. According to our best knowledge, no full-wave simulation with the field source carrying OAM has been published, so we compare the results from full-wave simulation in CST MWS with those from the numerical calculation based on the equivalent dipole model proposed in~\cite{menglin1}.

\begin{figure}[!t]
\centering
\includegraphics[width=\columnwidth]{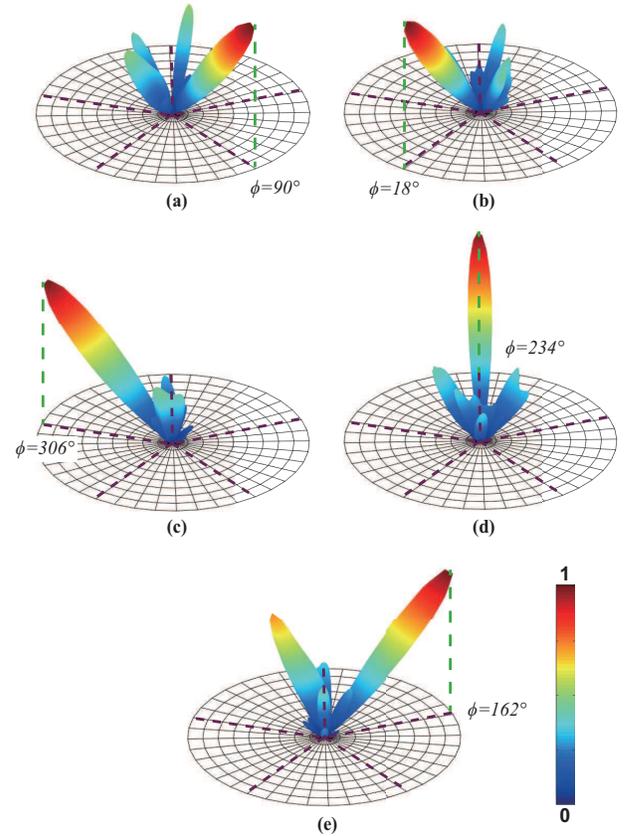}
\caption{Numerically calculated far-field power pattern when the incident wave carries OAM of order (a) -2; (b) -1; (c) 0; (d) 1; (e) 2.}
\label{radpat_dipole}
\end{figure}

\begin{figure}[!t]
\centering
\includegraphics[width=\columnwidth]{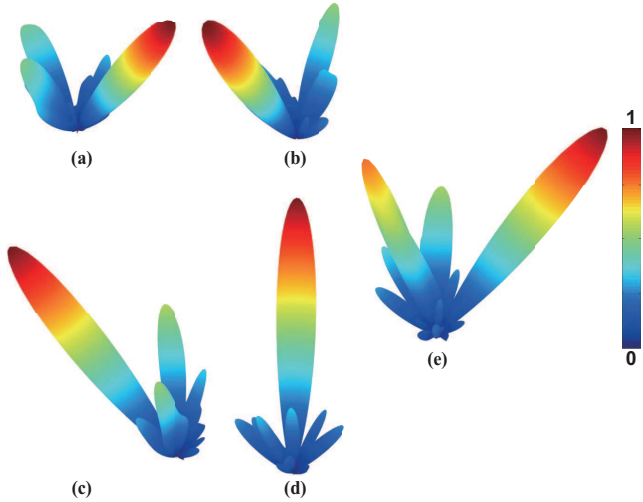}
\caption{Full-wave simulated far-field power pattern when the incident wave carries OAM of order (a) -2; (b) -1; (c) 0; (d) 1; (e) 2.}
\label{radpat_cst}
\end{figure}

The cross-circularly polarized components are examined. The numerical results based on the equivalent dipole model and full-wave simulated results are shown in Fig.~\ref{radpat_dipole} and Fig.~\ref{radpat_cst}, respectively. Whatever value of OAM the incident wave carries, the numerical and simulated radiation patterns show a good agreement with each other, which validates the correctness of full-wave simulation by importing the transverse OAM-source components in CST MWS. Meanwhile, the azimuthal locations of the main lobes under the five incident OAM waves are labeled in Fig.~\ref{radpat_dipole}(a) to (e), namely $\phi=90^\circ, 18^\circ, 306^\circ, 234^\circ, 162^\circ$, which is consistent with the theory.

It is worth noting that when the incidence wave carries higher order OAM, the side-lobe level increases. Particularly, for the case when $l_0=2$, the side lobe is very high. The high side lobe results from the constructive interference among multiple beams, especially between two adjacent beams. In Fig.~\ref{matlab_field}(c) and (d), there is a constructive interference between the beams with $l=2$ and $l=1$. The phase values of the two beams at the interference area are close to each other so that the two beams merge together. It is known that when two beams are superimposed, the resultant interference pattern depends on their relative amplitude and phase~\cite{composite}. Therefore, the constructive interference can be weakened by introducing different phase constants in the two beams. This phase constant, denoted by $\alpha_m$ is added into the transmittance function as below, 
\begin{equation}
t_{mod}(r,\phi) = \sum_{m} e^{j(l_m\phi+k_{xm}x+k_{ym}y+\alpha_m)}.
\end{equation}

\begin{figure}[!t]
\centering
\includegraphics[width=\columnwidth]{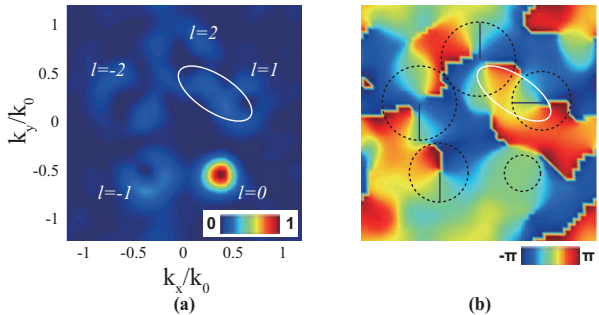}
\caption{Responds of the modified metasurface ($\alpha_2=-1.5708$) under the Gaussian beam incidence. The far-field (a) intensity and (b) phase patterns of the diffracted beam.}
\label{matlab_field1}
\end{figure}

To disturb the constructive interference in Fig.~\ref{matlab_field}(c) and (d), we can rotate one beam by adding a phase constant to it. A specific example is shown in Fig.~\ref{matlab_field1}. The beam with $l=1$ is rotated $-90^\circ$ by setting $\alpha_2=-1.5708$. This rotation changes the phase of the beam at the previous interference area, resulting in a weakened constructive interference.

\begin{figure}[!t]
\centering
\includegraphics[width=\columnwidth]{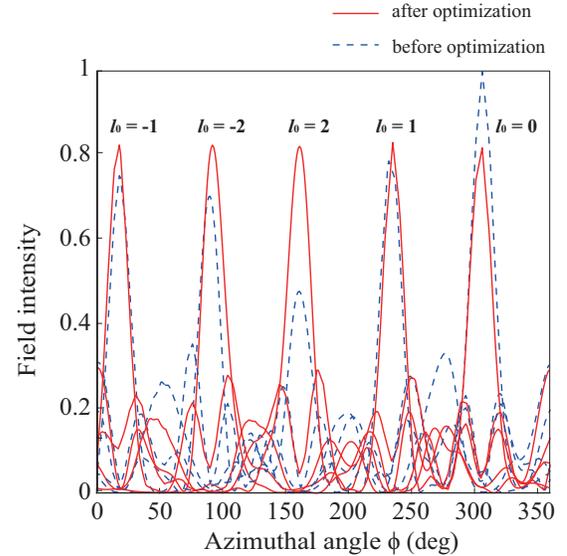}
\caption{Original and optimized far-field power patterns at $\theta=40^\circ$ for the five incident OAM waves.}
\label{radpat_cst_opt}
\end{figure}

Based on our analysis, the side lobes in Fig.~\ref{radpat_dipole} and Fig.~\ref{radpat_cst} can be suppressed by eliminating the constructive interference between any adjacent beams using the additional phase term $e^{i\alpha_m}$ in Eq.~(5). We use the brute-force method to find the values of $a_m$. For the five cases when the incident OAM $l_0=-2, -1, 0, 1$ and $2$, the side-lobe levels are different. So, our objective becomes minimizing the maximum peak-to-sidelobe ratio. The optimal solution is $\alpha_1=1.0472, \alpha_2=1.0472, \alpha_3=2.0944, \alpha_4=2.7925, \alpha_5=4.5379$. The comparison results after optimization are shown in Fig.~\ref{radpat_cst_opt}. The field intensity along the azimuthal coordinate is plotted at $\theta=40^\circ$ for each incident case. We can see that when $l_0=\pm1,\pm2$, the field intensity at desired location is increased. Particularly, for $l_0=2$, the field is enhanced significantly. Originally, the side lobe is the highest when $l_0=2$ and since our objective is to minimize the maximum peak-to-sidelobe ratio, it is reasonable that this scenario presents the most obvious improvement. Meanwhile, it can be noted that the intensity becomes lower for $l_0=0$. This is also a consequence of the objective we use for optimization. It is not possible to achieve improvement for all the five incident cases, but there has to be a trade off among each case to get the minimized maximum peak-to-sidelobe ratio. Overall, we can observe suppressed side lobes and increased field intensities at desired locations.

\section{Conclusion}
We proposed a novel approach for multiple OAM-beam detection by using a single metasurface. By examining the field intensity at the pre-designed locations, the OAM of the incident beam can be conveniently determined. A metasurface that can detect five OAM beams was demonstrated. Furthermore, based on the modified transmittance function, the metasurface was optimized for better performance. The effectiveness of our proposed scheme was verified by both numerical and full-wave simulations.

\section*{Acknowledgment}
This work was supported in part by the Research Grants Council of Hong Kong (GRF 716713, GRF 17207114, and GRF 17210815), NSFC 61271158, Hong Kong UGC AoE/P–04/08, AOARD FA2386-17-1-0010, Hong Kong ITP/045/14LP, and Hundred Talents Program of Zhejiang University (No. 188020*194231701/208).

\ifCLASSOPTIONcaptionsoff
  \newpage
\fi



%

\end{document}